\documentclass[twocolumn,aps,prl,floatfix,10pt,groupedaddress,longbibliography]{revtex4-1}
{}
\usepackage{pgfplots}
\usepackage{xcolor}
\usepackage{stackrel}
\usepackage[utf8]{inputenc} 
\usepackage{bm}
\usepackage{amsmath,amsthm,amssymb}
\usepackage{graphicx} 
\usepackage{siunitx}
\usepackage{tikz,tkz-tab}
\usepackage[
  colorlinks=true,
  urlcolor=blue,
  linkcolor=blue,
  citecolor=blue
]{hyperref}
\usepackage[normalem]{ulem}
\newcommand{\bs}{\boldsymbol}
\newcommand{\D}{{\rm d}}

 \newcommand{\I}{{\rm i}}
 \newcommand{\E}{{\rm e}}

 \newcommand{\kB}{k_{\rm B}}

\definecolor{red1}{RGB}{190,20,25}
\definecolor{red2}{RGB}{255,75,70}
\definecolor{red3}{RGB}{255,130,120}

\definecolor{blue1}{RGB}{25,100,170}
\definecolor{blue2}{RGB}{75,150,200}
\definecolor{blue3}{RGB}{150,200,220}

\definecolor{orange1}{RGB}{200,120,30}
\definecolor{orange2}{RGB}{255,170,100}

\definecolor{green1}{RGB}{25,125,75}
\definecolor{green2}{RGB}{120,200,125}

\definecolor{violet1}{RGB}{100,70,150}
\definecolor{violet2}{RGB}{150,150,200}

\definecolor{turquoise1}{RGB}{0,160,210}
\definecolor{turquoise2}{RGB}{115,210,230}

\definecolor{rose1}{RGB}{200,0,80}
\definecolor{rose2}{RGB}{230,120,165}

\definecolor{grey1}{RGB}{60,60,60}
\definecolor{grey2}{RGB}{180,180,180}

\begin{document}

\title{Tan's two-body contact across the superfluid transition of a planar Bose gas}

\author{Y.-Q. Zou}
\author{B. Bakkali-Hassani}
\author{C. Maury}
\author{\'E. Le Cerf}
\author{S. Nascimbene}
\author{J. Dalibard}
\author{J. Beugnon}

\affiliation{Laboratoire Kastler Brossel, Coll{\`e}ge de France, CNRS, ENS-PSL Research University, Sorbonne Universit{\'e}, 11 Place Marcelin Berthelot, 75005 Paris, France}

\date{\today}
\pacs{}

\begin{abstract}
Tan's contact is a quantity that unifies many different properties of a low-temperature gas with short-range interactions, from its momentum distribution to its spatial two-body correlation function. Here, we use a Ramsey interferometric method to realize experimentally the thermodynamic definition of the two-body contact, \emph{i.e.} the change of the internal energy in a small modification of the scattering length. Our measurements are performed on a uniform two-dimensional Bose gas of $^{87}$Rb atoms across the Berezinskii--Kosterlitz--Thouless superfluid transition. They connect well to the theoretical predictions in the limiting cases of a strongly degenerate fluid and of a normal gas. They also provide the variation of this key quantity in the critical region, where further theoretical efforts are needed to account for our findings.
\end{abstract}

\maketitle


The thermodynamic equilibrium of any homogeneous fluid is characterized by its equation of state. This equation gives the variations of a thermodynamic potential, \emph{e.g.} the internal energy $E$, with respect to a set of thermodynamics variables such as the number of particles, temperature, size and interaction potential. All items in this list are mere real numbers, except for the interaction potential whose characterization may require a large number of independent variables, making the determination of a generic equation of state challenging.

A considerable simplification occurs for ultra-cold atomic fluids, when the average distance between particles $d$ is much larger than the range of the potential between two atoms. Binary interactions can then be described by a single number, the s-wave scattering length $a$. Considering $a$ as a thermodynamic variable, one can define its thermodynamic conjugate, the so-called Tan's contact \cite{tan2008large,baym2007coherence,punk2007theory,braaten_exact_2008,werner_number_2009,Zhang_Leggett:2009_PhysRevA.79.023601,Combescot:2009_PhysRevA.79.053640,haussmann2009spectral,Braaten:2010}
\begin{equation}
C\equiv \frac{8\pi m a^2 }{\hbar^2} \; \frac{\partial E}{\partial a},
\label{eq:contact_definition}
\end{equation}
where the derivative is taken at constant atom number, volume and entropy, and $m$ is the mass of an atom. For a pseudo-spin 1/2 Fermi gas with zero-range interactions, one can show that the conjugate pair $(a,C)$ is sufficient to account for all possible regimes for the gas, including the strongly interacting case $a\gtrsim d$ \cite{petrov_three-body_2003,Endo:2015_PhysRevA.92.053624}. For a Bose gas, the situation is  more complicated: formally, one  needs to introduce also a parameter related to three-body interactions, and in practice this three-body contact can play a significant role in the strongly interacting regime \cite{Braaten:2011_PhysRevLett.106.153005,werner_general_2012Bosons,Smith:2014_PhysRevLett.112.110402}. 

Since the pioneering experimental works of \cite{stewart2010verification,Kuhnle:2010_PhysRevLett.105.070402}, the two-body contact has been used to relate numerous measurable quantities regarding interacting Fermi gases:  tail of the momentum distribution, short distance behavior of the two-body correlation function, radio-frequency spectrum in a magnetic resonance experiment, etc.\; (see \cite{carcy_contact_2019,Mukherjee:2019_PhysRevLett.122.203402} and refs.\;in). For the Bose gas case of interest here, experimental determinations of two- and three-body contacts are much more scarce, and concentrated so far on either the quasi-pure BEC regime \cite{wild2012measurements,Lopes:2017_PhysRevLett.118.210401} or the thermal one \cite{wild2012measurements,Fletcher:2017}. Here, we use a two-pulse Ramsey interferometric scheme to map out the variations of the two-body contact from the strongly degenerate, superfluid case to the non-degenerate, normal one.

We operate with a uniform, weakly-interacting two-dimensional (2D) Bose gas where the superfluid transition is of Berezinskii--Kosterlitz--Thouless (BKT) type \cite{Berezinskii:1971,Kosterlitz:1973}.  For our relatively low spatial density, effects related to the three-body contact are negligible and we focus on the two-body contact. It is well known that for the BKT transition, all thermodynamic functions are continuous at the critical point, except for the superfluid density \cite{Kosterlitz17}. Our measurements confirm that the two-body contact is indeed continuous at this point. We also show that the (approximate) scale invariance in 2D allows us to express it as a function of a single parameter, the phase-space density ${\cal D}=n\lambda^2$, where $n$ is the 2D density, $\lambda=(2\pi \hbar^2/mk_{\rm B}T)^{1/2}$ the thermal wavelength and $T$ the temperature. Our measurements around the critical point of the BKT transition provides an experimental milestone which shows the limits of the existing theoretical predictions in the critical region.

\section{Results}

\paragraph{Accessing Tan's contact for a planar geometry.} 

Our ultra-cold Bose gas is well  described by the Hamiltonian $\hat H$, sum of the kinetic energy operator, 
the confining potential,  
and the interaction potential $\hat H_{\rm int}=a \hat K$ with
\begin{equation}
\hat K=\frac{2\pi \hbar^2}{m} \int \hskip -3mm\int  \hat \psi^\dagger(\bs r)\, \hat \psi^\dagger(\bs r')\;\hat \delta(\bs r-\bs r')\;\hat \psi(\bs r')\,\hat \psi(\bs r)\ \D^3 r\;\D^3 r'.
\label{eq:interaction_potential_K}
\end{equation}
Here $\hat \delta (\bs r)$ is the regularized Dirac function entering in the definition of the pseudo-potential \cite{huan87} and the field operator $\hat \psi(\bs r)$ annihilates a particle in $\bs r$. Using Hellmann--Feynman theorem, one can rewrite the contact defined in Eq.\,(\ref{eq:contact_definition}) as 
$C=8\pi m a^2 \langle \hat K\rangle/\hbar^2$. 

In our experiment, the gas is uniform in the horizontal $xy$ plane, and it is confined with a harmonic potential of frequency $\omega_z$ along the vertical direction. We choose $\hbar \omega_z$ larger than both the interaction energy and the temperature, so that the gas is thermodynamically two-dimensional (2D). On the other hand, the extension of the gas $a_z=(\hbar /m\omega_z)^{1/2}$ along the  direction $z$ is still large compared to the scattering length $a$, so that the collisions  keep their 3D character and Eq.\,(\ref{eq:interaction_potential_K}) remains relevant  \cite{Petrov:2001}. Suppose first that the zero-range potential $\hat \delta(\bs r-\bs r')$ appearing in (\ref{eq:interaction_potential_K}) does not need to be regularized. Then, after integration over $z$, $C$ can be related to the in-plane two-body correlation function $g_2$:
\begin{equation}
\frac{C}{C_0}\stackbin{?}{=} g_2(0) , \qquad C_0\equiv 4(2\pi)^{3/2} \frac{a^2 \bar n N}{a_z},
\label{eq:relation_C_g2}
\end{equation}
where we introduced the average in normal order:
\begin{equation}
g_2(\bs r)=\frac{1}{\bar n^2}\langle :\hat n(\bs r) \hat n(0):\rangle,
\end{equation}
with $\hat n(\bs r)$ the operator associated with the 2D density, $\bar n$ its average value and $N$ the atom number. For an ideal Bose gas, the value of $g_2(0)$ varies from $2$  to 1 when one goes from the non-condensed regime to the fully condensed one \cite{Naraschewski:1999_PhysRevA.59.4595}, so that $C_0$ sets the scale of Tan's contact.

However, it is well known that $g_2(0)$ is generally an ill-defined quantity for an interacting fluid. For example in a Bose gas with zero-range interactions, one expects $g_2(r)$ to diverge as $1/r^2$ in 3D and $(\log r)^2$ in 2D when $r\to 0$
\cite{Braaten:2011_PhysRevLett.106.153005,werner_general_2012Bosons}. On the other hand, when one properly regularizes the zero-range potential $\hat \delta$ in Eq.\,(\ref{eq:interaction_potential_K}), Tan's contact is well-behaved and measurable. Here, we approach it by measuring the change in energy per atom $h \Delta \nu=\Delta E/N$ when the scattering length is changed by the small amount $\Delta a$. Replacing $\partial E/\partial a$ by $\Delta E/\Delta a$ in the definition (\ref{eq:contact_definition}), we obtain
\begin{equation}
\frac{C}{C_0}\approx \sqrt{2\pi} \;\frac{m a_z}{\hbar \bar n}\;\frac{\Delta \nu}{\Delta a}.
\label{eq:C_C0_Delta_nu}
\end{equation}

To measure the energy change $h\Delta \nu$ resulting for a small modification of the scattering length, we take advantage of a particular feature of the $^{87}$Rb atom: All scattering lengths $a_{ij}$, with $(i,j)$ any pair of states belonging to the ground-level manifold, take very similar values  \cite{vanKempen:2002_PhysRevLett.88.093201}. For example, Ref.\,\cite{altin2011optically} predicts $a_{11}=100.9\,a_0$, $a_{22}=94.9\,a_0$ and $a_{12}=98.9\,a_0$, where the indices 1 and 2 refer to the two states $|1\rangle\equiv |F=1,m_z=0\rangle$ and $|2\rangle\equiv |F=2,m_z=0\rangle$ used in this work and $a_0$ is the  Bohr  radius. For an isolated atom, this pair of states forms the so-called clock transition at frequency $\nu_0\simeq 6.8\,$GHz, which is insensitive (at first order) to the ambiant magnetic field. Starting from a gas at equilibrium in $|1\rangle$, we use a Ramsey interferometric scheme to measure the microwave frequency required to transfer all atoms to the state $|2\rangle$. The displacement of this frequency with respect to $\nu_0$ provides the shift $\Delta \nu$ due to the small modification of scattering length $\Delta a=a_{22}- a_{11}$.


\begin{figure}[t]
  \begin{center}
    \includegraphics{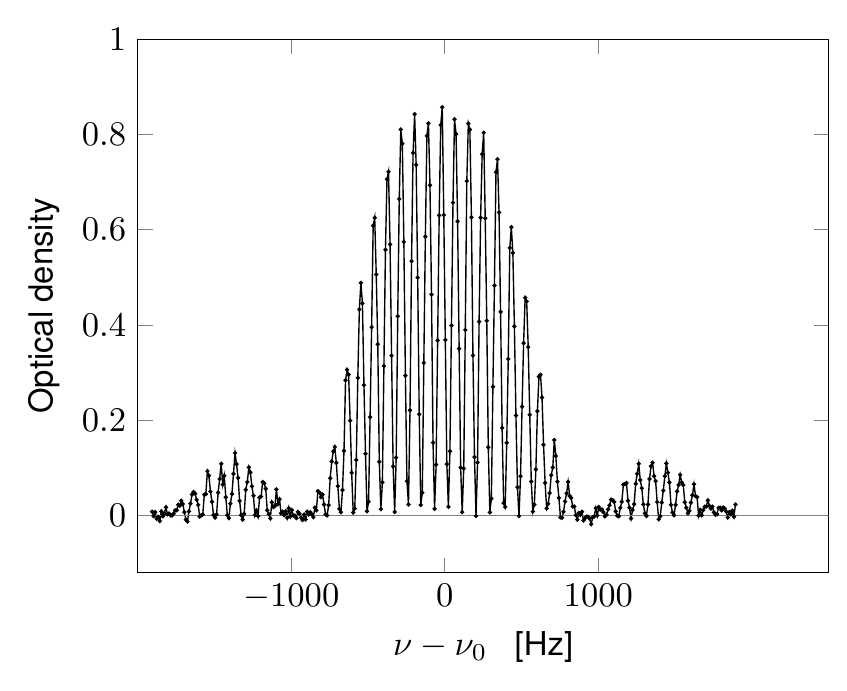}
      \vskip -67.8mm
      \hskip 56mm
    \includegraphics[width=30mm]{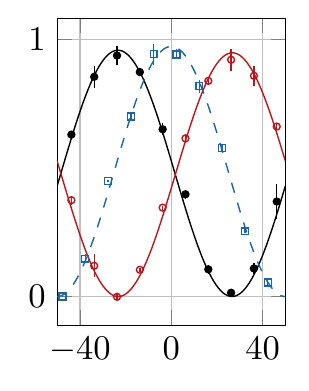}
      \vskip 26mm
  \end{center}
  \caption{Example of an interferometric Ramsey signal showing the optical density of the fraction of the gas in state $|2\rangle$ after the second Ramsey pulse, as a function of the microwave frequency $\nu$. These data were recorded for $\bar n\approx 40$ atoms/\si{\micro}m$^2$ and $T\sim 22\,$nK, $\tau_1=10\,$ms. Here, $\tau_2$ has been increased to 1\,ms to limit the number of fringes for a better visibility. Inset. Filled black disks (resp. open red circles): central fringe for atoms in $|2\rangle$ (resp. $|1\rangle$) in the ``standard" configuration $\tau_2=0.1\,$ms. The density in $|1\rangle$ is obtained by applying a  microwave $\pi$-pulse just before the absorption imaging phase. Blue squares: single-atom response measured during the ballistic expansion of the cloud by imaging atoms in $|2\rangle$. The lines in the inset are sinusoidal fits to the data. The vertical error bars of the inset correspond to the standard deviation of the 3 repetitions made for this measurement.}
  \label{fig:Ramsey_signal}
\end{figure}


\paragraph{Ramsey spectroscopy on the clock transition.} 

The Ramsey scheme consists in two identical microwave pulses, separated by a duration $\tau_1 = 10\,$ms. Their duration $\tau_2\sim 100\,$\si{\micro}s is adjusted to have $\pi/2$ pulses, \emph{i.e.}\;each pulse brings an atom initially in $|1\rangle$ or $|2\rangle$ into a coherent superposition  of these two states with equal weights. Just after the second Ramsey pulse, we measure the 2D spatial density $\bar n$ in state $|2\rangle$ in a disk-shaped region of radius 9\,\si{\micro}m  and using the absorption of a probe beam nearly resonant with the optical transition connecting $|2\rangle$ to the excited state $5P_{3/2},\, F'=3$. We infer from this measurement the fraction of atoms transferred into $|2\rangle$ by the Ramsey sequence, and we look for the microwave frequency $\nu_m$ that maximises this fraction. 

An example of spectroscopic signal is shown in Fig. \ref{fig:Ramsey_signal}. In order to determine the ``bare" transition frequency $\nu_0$, we also perform a similar measurement on a cloud in ballistic expansion, for which the 3D spatial density has been divided by more than 100 and interactions play a negligible role. The uncertainty on the measured interaction-induced shift $\Delta \nu=\nu_m-\nu_0$ is on the order of 1 Hz. In principle, the precision of our measurements could be increased further by using a larger $\tau_1$. In practice however, we have to restrict $\tau_1$ to a value such that the spatial dynamics of the cloud, originating from the non-miscibility of the $1-2$ mixture ($a_{12}^2>a_{11} a_{22}$), plays a negligible role \footnote{We also check that no detectable spin-changing collisions appear on this time scale: more than 99\,\% of the atoms stay in the clock state basis.}. Another limitation to $\tau_1$ comes from atom losses, mostly due to 2-body inelastic processes involving atoms in $|2\rangle$. For $\tau_1=10\,$ms, these losses affect less than $5\%$ of the total population and can be safely neglected. 

We see in Fig.\,\ref{fig:Ramsey_signal} that there indeed exists a frequency $\nu_m$ for which nearly all atoms are transferred from $|1\rangle$ to $|2\rangle$, so that $E(N,a_{22})-E(N,a_{11})=N\,h(\nu_m-\nu_0)$ (see \cite{SM} for details). We note that for an interacting system, the existence of such a frequency is by no means to be taken for granted. Here, it is made possible by the fact that the inter-species scattering length $a_{12}$ is close to $a_{11}$ and $a_{22}$. We are thus close to the SU(2) symmetry point where all three scattering lengths coincide. The modeling of the Ramsey process detailed in \cite{SM} shows that this quasi-coincidence allows one to perform a Taylor expansion of the energy $E(N_1,N_2)$ (with $N_1+N_2=N$) of the mixed system between the two Ramsey pulses, and to expect a quasi-complete rephasing of the contributions of all possible couples $(N_1,N_2)$ for the second Ramsey pulse. The present situation is thus quite different from the one exploited in 
\cite{Fletcher:2017} for example, where $a_{11}$ and $a_{12}$ were vanishingly small. It also differs from the generic situation prevailing in the spectroscopic measurements of Tan's contact in two-component Fermi gases, where a microwave pulse transfers the atoms to a third, non-interacting state \cite{stewart2010verification}.  


\begin{figure}[t]
  \begin{center}
    \includegraphics{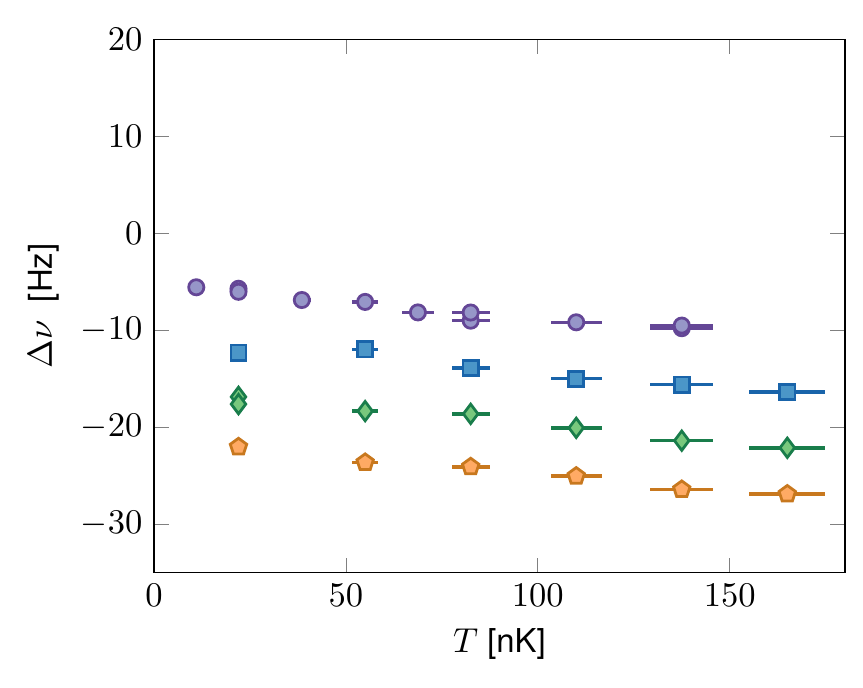}
      \vskip -68.5mm
      \hskip 44mm
    \includegraphics{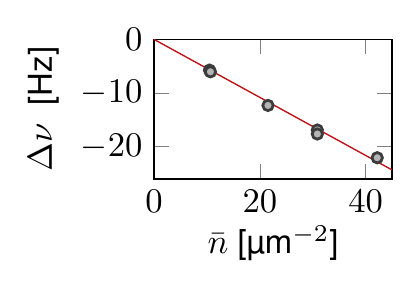}
      \vskip 36mm 
  \end{center}
  \caption{Variations of the shift $\Delta \nu$ with temperature for various 2D spatial densities. Violet disks: $\bar n=10.4\,(2)$\,\si{\micro}m$^{-2}$, blue squares:  $\bar n=21.0\,(3)$\,\si{\micro}m$^{-2}$, green diamonds: $\bar n=31.5\,(3)$\,\si{\micro}m$^{-2}$, orange pentagons: $\bar n=42.0\,(1)$\,\si{\micro}m$^{-2}$. The horizontal error bars represent the statistical uncertainty on the temperature calibration, except for the points at very low temperature (10-22\,nK). These ultracold points are deeply in the Thomas-Fermi regime, where thermometry based on the known equation of state of the gas is not sensitive enough. The temperature is thus inferred from an extrapolation with evaporation barrier height of the higher temperature points. The error on the frequency measurement is below 1\,Hz and is not shown in this graph. Inset: Variations of the shift $\Delta \nu$ with density at low temperature $T \sim 22$\,nK, \emph{i.e.}\;a strongly degenerate gas. The straight line is the mean-field prediction corresponding to $\Delta a=-5.7\,a_0$.}
  \label{fig:Delta_nu_vs_T}
\end{figure}


\paragraph{Resonance shift $\Delta \nu$ and contact $C$.}

We show in Fig.\,\ref{fig:Delta_nu_vs_T} our measurements of the shift $\Delta \nu$ for  densities ranging from 10 to  40 atoms/\si{\micro}m$^2$, and temperatures from 10 to 170\,nK. Since $\hbar \omega_z/k_{\rm B}=210\,$nK, all data shown here are in the thermodynamic 2D regime $k_{\rm B}T<\hbar \omega_z$. More precisely, the population of the ground state of the motion along $z$, estimated from the ideal Bose gas model  \cite{Chomaz:2015}, is always $\gtrsim$ 90\,\%. All shifts are negative as a consequence of $a_{22}<a_{11}$: the interaction energy of the gas in state $|2\rangle$ is slightly lower than in state $|1\rangle$. For a given density, the measured shift increases in absolute value with temperature. This is in line with the naive prediction of Eq.\,(\ref{eq:relation_C_g2}), since density fluctuations are expected to be an increasing function of $T$. Conversely for a given temperature, the shift is (in absolute value) an increasing function of density.

For the lowest temperatures investigated here, we reach the fully condensed regime in spite of the 2D character of the sample, as a result of finite size effects. In this case, the mean-field prediction for the shift reads $\Delta \nu=\bar n \, \hbar\, \Delta a/(\sqrt{2\pi}\, m a_z)$ [\emph{i.e.}\;$C=C_0$ in Eq.\,(\ref{eq:C_C0_Delta_nu})]. Our measurements confirm the linear variation of $\Delta \nu$ with $\bar n$, as shown in the inset of Fig.\,\ref{fig:Delta_nu_vs_T} summarizing the data for $T=22\,$nK. A linear fit to these data gives $\Delta a/a_0=-5.7\,(1.0)$ where the error mostly originates from the uncertainty on the density calibration. In the following, we use this value of $\Delta a$ for inferring the value of $C/C_0$ from the measured shift at any temperature, using Eq.\,(\ref{eq:C_C0_Delta_nu}). We note that this estimate for $\Delta a$ is in good agreement with the prediction $\Delta a/a_0=-6$ quoted in \cite{altin2011optically}, as well as with our recent measurement \cite{Zou2020} which is independent of the density calibration. The first corrections to the linear mean-field prediction were derived (in the 3D case) by Lee, Huang and Yang in \cite{Lee:1957}. For our densities, they have a relative contribution on the order of 5\,\% of the main signal ($\Delta \nu \lesssim 1\,$Hz) \cite{SM}, and their detection is borderline for our current precision.  

\begin{figure}[t]
  \begin{center}
    \includegraphics{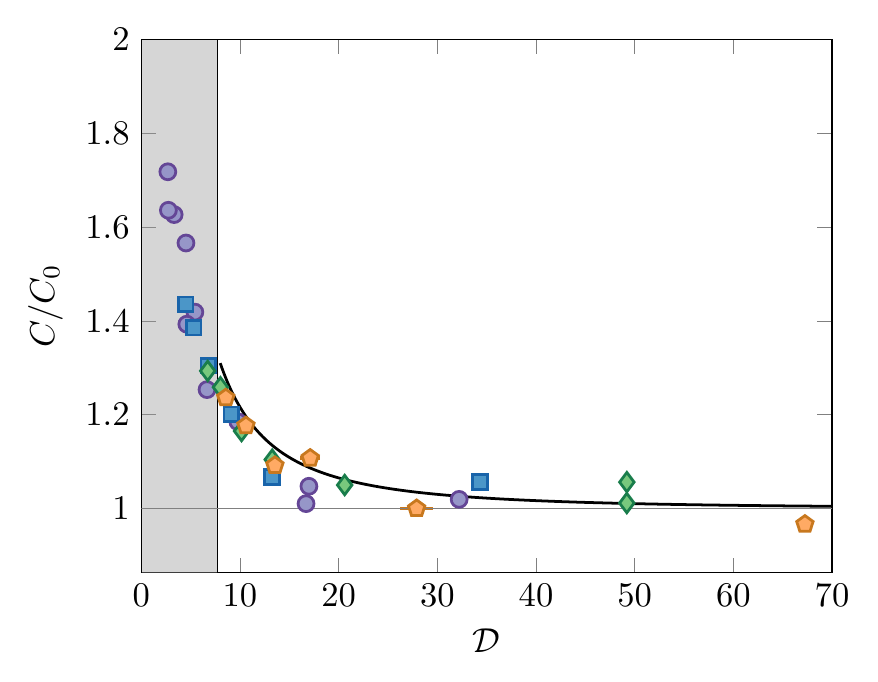}
      \vskip -65mm
      \hskip 30mm
    \includegraphics{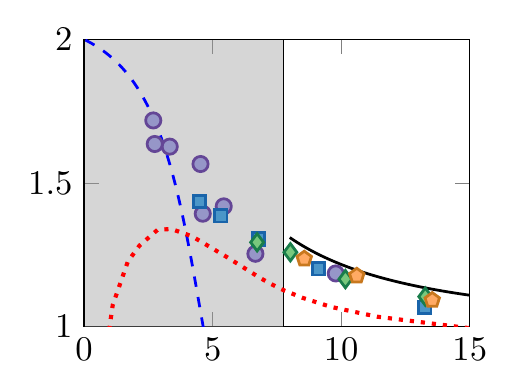}
      \vskip 23mm 
  \end{center}
  \caption{Variations of the normalized Tan's contact $C/C_0$ with the phase-space density ${\cal D}$. The encoding of the experimental points is the same as in Fig. \ref{fig:Delta_nu_vs_T}. The colored zone indicates the non-superfluid region, corresponding to  ${\cal D}<{\cal D}_{\rm c}\approx 7.7$. The continuous black line shows the prediction derived within Bogoliubov approximation. Inset: Zoom on the critical region. The dashed blue line is the prediction from \cite{ren2004virial}, resulting from a virial expansion for the 2D Bose gas. The dotted red line shows the results of the classical field simulation of \cite{Prokofev:2002}.}
  \label{fig:Contact_vs_TTc}
\end{figure}

We  summarize all our data in Fig.\,\ref{fig:Contact_vs_TTc}, where we show the normalized contact $C/C_0$ defined in Eq.\,(\ref{eq:C_C0_Delta_nu}) as a function of the  phase-space density ${\cal D}$. All data points collapse on a single curve within the experimental error, which is a manifestation of the approximate scale invariance of the Bose gas, valid for a relatively weak interaction strength $\tilde g\lesssim 1$ \cite{Hung:2011,Yefsah:2011}. 


\section{Discussion}
We now compare our results  in Fig.\,\ref{fig:Contact_vs_TTc} to three theoretical predictions. The first one is derived from the Bogoliubov approximation applied to a 2D quasi-condensate \cite{Mora:2003}. This prediction is expected to be valid only for ${\cal D}$ notably larger than  the phase-space density at the critical point ${\cal D}_c$ (see methods), but it gives a fair account of our data over the whole superfluid region. Within this approximation, one can also calculate the two-body correlation function and write it as  $g_2(r)= g_2^{T=0}(r)+g_2^{\rm thermal}(r)$. One can then show the result \cite{SM}
\begin{equation}
\frac{C}{C_0}=1+g_2^{\rm thermal}(0),
\end{equation}
which provides a quantitative relation between the contact and the pair correlation function, in spite of the already mentioned singularity of $g_2^{T=0}(r)$ in $r=0$.

For low phase-space densities, one can perform a systematic expansion of various thermodynamic functions in powers of the (properly renormalized) interaction strength  \cite{ren2004virial}, and obtain a prediction for $C$ (dashed blue line in the inset of Fig.\,\ref{fig:Contact_vs_TTc}). By comparing the 0th, 1st and 2nd orders of this virial-type expansion, one can estimate that it is valid for ${\cal D}\lesssim 3$ for our parameters. When ${\cal D}\to 0$, the result of \cite{ren2004virial} gives $C/C_0\to 2$, which is the expected result for an ideal, non-degenerate Bose gas. The prediction of   \cite{ren2004virial} for ${\cal D}\sim 3$ compares favourably with our results in the weakly-degenerate case. 

Finally we also show in Fig.\,\ref{fig:Contact_vs_TTc} the results of the classical field simulation of \cite{Prokofev:2002} (red dotted line), which are  in principle valid both below and above the critical point. Contrary to the quantum case, this classical analysis does not lead to any singularity for $\langle n^2(0)\rangle$, so that we can directly plot this quantity as it is provided in \cite{Prokofev:2002} in terms of the quasi-condensate density. For our interaction strength, we obtain a non-monotonic variation of $C$. This unexpected behavior, which does not match the experimental observations, probably signals that the present interaction strength $\tilde g=0.16$ (see Methods) is too large for using these classical field predictions, as already suggested in \cite{Prokofev:2002}.


Using the Ramsey interferometric scheme on a many-body system, we have measured the two-body contact of a 2D Bose gas over a wide range of phase-space densities. We could implement this scheme on our fluid thanks to the similarities of the three scattering lengths in play, $a_{11},a_{22},a_{12}$, corresponding to an approximate SU(2) symmetry for interactions. Our method  can be generalized to the strongly interacting case $a_{ij}\gtrsim a_z$, as long as a Fano-Feshbach resonance allows one to stay close to the SU(2) point. One could then address simultaneously the LHY-type corrections at zero temperature \cite{Mora:2009_PhysRevLett.102.180404,Fournais:2019}, the contribution of the three-body contact \cite{werner_general_2012Bosons,Smith:2014_PhysRevLett.112.110402},
and the breaking of scale invariance expected at non-zero temperature. Finally we note that even for our moderate interaction strength, classical field simulations seem to fail to reproduce our results, although they could properly account for the measurement of the equation of state itself \cite{Hung:2011,Yefsah:2011}. The semi-classical treatment of Ref.\,\cite{Giorgetii:2007_PhysRevA.76.013613} and quantum Monte Carlo approaches of Refs.\,\cite{Holzmann08,Rancon12} should provide a reliable path to the modelling of this system. This would be particularly interesting in the vicinity of the BKT transition point where the usual approach based on the $XY$ model \cite{Nelson:1977}, which neglects any density fluctuation, does not provide a relevant information on the behavior of Tan's contact.


\section{Methods}

\paragraph{Preparation of the two-dimensional gas.}
 
The preparation and the characterization of our sample have been detailed in  \cite{Ville:2017,Ville:2018} and we briefly outline the main properties of the  clouds explored in this work. In the $xy$ plane, the atoms are confined in a disk of radius $12\,$\si{\micro}m by a box-like potential, created by a laser beam properly shaped with a digital micromirror device. We use the intensity of this beam, which determines the height of the potential barrier around the disk, as a control parameter for the temperature. The confinement along the $z$ direction is provided by a large-period optical lattice, with a single node  occupied and $\omega_z/(2\pi)= 4.41\,(1)\,$kHz. We set a magnetic field $B=0.701\,(1)$\,G along the vertical direction $z$, which defines the quantization axis.
We use the expression ${\cal D}_{\rm c}=\ln(380/ \tilde g)$ for the phase-space density at the critical point of the superfluid transition \cite{Prokofev:2001}. Here, $\tilde g=\sqrt{8\pi}\,a_{11}/a_z=0.16$ is the dimensionless interaction strength in 2D, leading to ${\cal D}_{\rm c}=7.7$. We study Bose gases from the normal regime (${\cal D}=0.3 {\cal D}_{\rm c}$) to the strongly degenerate, superfluid regime (${\cal D}>3 {\cal D}_{\rm c}$).

\paragraph{Acknowledgments.}	
We thank Paul Julienne, Raphael Lopes, and F\'elix Werner for useful discussions. We acknowledge the contribution of Rapha\"el Saint-Jalm at the early stage of the project. This work was supported by ERC (Synergy Grant UQUAM), Quantera ERA-NET (NAQUAS project) and the ANR-18-CE30-0010 grant. LKB is a member of the SIRTEQ network of R\'egion Ile-de-France. 

\paragraph{Author contributions.} 
Y.-Q.Z., B.B.-H. and C.M. performed the experiment and carried out the preliminary data analysis. Y.-Q.Z. performed the detailed data analysis. E.L.C. participated in the preparation of the experimental setup. S.N., J.D. and J.B. contributed to the development of the theoretical model.  J.D. and J.B. wrote the manuscript with contributions from all authors.

\appendix

\section{Supplementary Material}

\section{Ramsey interferometry in a many-body system close the SU(2) symmetry point}

In this section, we explain why the vicinity of the SU(2) symmetry point where all three scattering lengths are equal ($a_{11}=a_{12}=a_{22}$) allows one to reach a full transfer from $|1\rangle$ to $|2\rangle$ in the Ramsey sequence, in spite of the interactions between the particles. We first explore a two-particle model before turning to the general $N$-atom case.

\subsection{The two-particle toy model}

\begin{figure}[t]
\begin{center}
\includegraphics{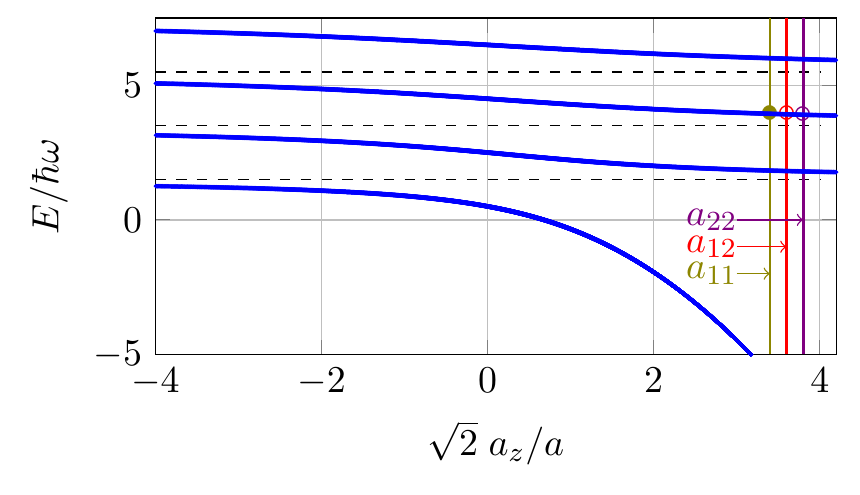}
\end{center}
\caption{Energy levels of the relative motion of zero angular momentum for a two-particle system in a 3D harmonic trap of frequency $\omega$, as function of the scattering length. To model properly the experimental situation, the characteristic length $a_z=\sqrt{\hbar/m\omega}$ is chosen equal to the interparticle distance $d=\bar n^{-1/2}$ ($d\sim a_z\sim 160\,$nm for $\bar n=40\,$\si{\micro}m$^{-2}$, i.e. $\hbar \omega=\hbar^2 \bar n /m$. Therefore the spacing $\sim 2\hbar \omega$ between adjacent levels is large compared to the interaction energy per particle, 
$\sim \hbar^2 \bar n \tilde g /m$, since $\tilde g\ll 1$. 
The initial state  $|\psi_0\rangle$ considered in the text is marked as $\bullet$ and the two other relevant states $|\phi_0\rangle$ and $|\chi_0\rangle$ are marked as $\circ$. All three scattering lengths $a_{11}, a_{12},a_{22}$ are close to each other (figure not to scale for actual Rb values).}
\label{fig:Busch}
\end{figure}

The analysis of a system with two particles only, which was pioneered by \cite{Busch:1998}, is often used to gain insight in the $N$-body case, see e.g. \cite{sykes2014quenching,Fletcher:2017} in the context of microwave spectroscopy. Here we consider a pair of atoms each with two internal states $|1\rangle$ and $|2\rangle$ (Fig.\ref{fig:Busch}). The initial state of the two-particle system is 
\begin{equation}
|11\rangle \otimes |\psi_0\rangle,
\end{equation}
where $|\psi_0\rangle$ describes the external state of the pair and is symmetric by exchange of the two (bosonic) particles.

The two-body state just after the first $\pi/2$ pulse of the Ramsey sequence is
\begin{equation}
\left[\frac{1}{2}|A\rangle + \frac{1}{\sqrt 2}|B\rangle + \frac{1}{2}|C\rangle  \right] \otimes |\psi_0\rangle.
\end{equation}
Here we have introduced the three states 
\begin{equation}
|A\rangle=|11\rangle\qquad |B\rangle=\frac{1}{\sqrt 2}\left(|12\rangle + |21\rangle  \right) \qquad |C\rangle=|22\rangle
\end{equation}
which correspond to the triplet states, resulting from the coupling of the two internal states viewed as pseudo-spins $1/2$.

The time evolution is described by three operators $\hat U_{ij}(t)$ and the state of the system reads at time $t$:
\begin{eqnarray}
&& \frac{1}{2}|A\rangle\otimes \left(\hat U_{11}(t)|\psi_0\rangle\right)\ +\ \frac{1}{\sqrt 2}|B\rangle\otimes \left(\hat U_{12}(t)|\psi_0\rangle\right) \nonumber\\
 &+&\ \frac{1}{2}|C\rangle   \otimes \left(\hat U_{22}(t)|\psi_0\rangle\right).
\end{eqnarray}
The action of the second $\pi/2$ pulse at time $t$ reads:
\begin{equation}
|1\rangle \rightarrow \frac{1}{\sqrt 2}\left( |1\rangle +\E^{\I \alpha}|2\rangle\right) \quad
|2\rangle \rightarrow \frac{1}{\sqrt 2}\left( |2\rangle -\E^{-\I \alpha}|1\rangle\right),
\label{eq:second_pulse}
\end{equation}
where $\alpha=2\pi \nu t$ is the phase of the microwave at this time.
After the second pulse, we find the fraction $f_2(t)$ transferred to internal state $|2\rangle$:
\begin{equation}
f_2(t,\alpha) =\frac{1}{2}+\frac{1}{4}\Re \left[\E^{\I \alpha} \left( \langle \hat U_{12}^\dagger \hat U_{11}\rangle 
+ \langle \hat U_{22}^\dagger \hat U_{12}\rangle \right)  \right],
\label{eq:Ramsey_signal_2}
\end{equation}
where the averages are taken in state $|\psi_0\rangle$.

The contact is calculated as the derivative with respect to the scattering length of the energy of the system (here the pair of atoms) at constant entropy and in thermal equilibrium. Therefore we can suppose that $|\psi_0\rangle$ is an eigenstate of the two-particle system for the scattering length $a_{11}$ and eventually perform a statistical average over $|\psi_0\rangle$ at the end of the analysis. 
 
To calculate the various matrix elements $\langle \hat U_{ij}^\dagger \hat U_{kl}\rangle$ entering in the expression (\ref{eq:Ramsey_signal_2}) of the Ramsey signal, we introduce the eigenbases of the two-particle system for the scattering lengths $a_{12}$ and $a_{22}$,  denoted respectively 
$\{|\phi_n\rangle\}$ and $\{|\chi_n\rangle\}$. For $^{87}$Rb, the three scattering lengths  $a_{11}, a_{12},a_{22}$ are close to each other (5\% difference at most). This means that essentially one state contributes
to the expansion of $|\psi_0\rangle$ on the basis $\{|\phi_n\rangle\}$ or $\{|\chi_n\rangle\}$:
\begin{equation}
|\psi_0\rangle \approx |\phi_0\rangle \approx |\chi_0\rangle .
\end{equation}    
This validates the assumption of constant entropy needed for the calculation of the contact: the populations of the eigenstates of the external motion of the two-particle system are quasi-unchanged by the Ramsey pulses (Fig.\ref{fig:Busch}).

With this assumption, we find
\begin{equation}
\langle \hat U_{12}^\dagger \hat U_{11}\rangle \approx \E^{\I (E_{12}-E_{11})t/\hbar },\quad 
\langle \hat U_{22}^\dagger \hat U_{12}\rangle \approx \E^{\I (E_{22}-E_{12})t/\hbar }
\label{eq:matrix_elements}
\end{equation}
where $E_{ij}$ includes both the single atom energy $\pm h\nu_0/2$ and the interaction energy of the atom pair. The Ramsey signal now reads:
\begin{eqnarray}
f_2(t)&\approx& \frac{1}{2}+\frac{1}{4}\cos\left[\alpha+(E_{12}-E_{11})t/\hbar \right] \nonumber \\
&& \quad +\frac{1}{4}\cos\left[\alpha+(E_{22}-E_{12})t/\hbar \right] .
\label{eq:Ramsey_2_bis}
\end{eqnarray}
It is maximal for
\begin{equation}
2h\nu=E_{22}-E_{11} 
\label{eq:alpha_opt_2_body}
\end{equation}
as announced in the main text, and it reaches $f_2=1$ when $a_{12}$ is equal to the arithmetic mean of $a_{11}$ and $a_{22}$. These conclusions are unchanged when one subsequently performs a statistical average over $|\psi_0\rangle$. 


\subsection{Achieving a full transfer in the $N$-body Ramsey sequence}

We consider a collection of $N$ two-level atoms with internal states $|1\rangle,|2\rangle$, and we assume that the initial state for the Ramsey sequence corresponds to having all atoms in the internal state $|1\rangle$:
\begin{equation}
|\Psi_0\rangle=\frac{1}{\sqrt{N!}}\left(\hat a_1^\dagger\right)^N \;|0\rangle,
\end{equation}
with a given external many-body state $|\psi_0\rangle$. 
  
After the first $\pi/2$ pulse, the collective internal state is
\begin{eqnarray}
|\Psi_1\rangle&=&\frac{1}{\sqrt{2^N\,N!}}\left(\hat a_1^\dagger+\hat a_2^\dagger\right)^N |0\rangle
\nonumber \\
&=&\frac{1}{\sqrt{2^N\,N!}}\sum_{N_1=0}^N \binom{N}{N_1} \left(\hat a_1^\dagger\right)^{N_1} \left(\hat a_2^\dagger\right)^{N_2} \;|0\rangle.
\label{eq:binomial_expansion}
\end{eqnarray}
We denote $E(N_1,N_2)$ the energy of the system with $N_1$ particles in $|1\rangle$ and $N_2=N-N_1$ particles in $|2\rangle$. After the evolution for a duration $t$, the state becomes:
\begin{equation}
\frac{1}{\sqrt{2^N\,N!}}\sum_{N_1=0}^N \binom{N}{N_1}\;\E^{-\I E(N_1,N_2)t/\hbar} \left(\hat a_1^\dagger\right)^{N_1} \left(\hat a_2^\dagger\right)^{N_2} \;|0\rangle.
\label{eq:eq:binomial_expansion_t}
\end{equation}
The second $\pi/2$ pulse at time $t$ corresponds to 
\begin{equation}
\hat a_1^\dagger \to \frac{1}{\sqrt 2}\left(\hat a_1^\dagger +\E^{\I \alpha}\hat a_2^\dagger  \right), \quad
\hat a_2^\dagger \to \frac{1}{\sqrt 2}\left(\hat a_2^\dagger -\E^{-\I \alpha}\hat a_1^\dagger  \right),
\end{equation}
where $\alpha=2\pi \nu t$ is the phase of the microwave at time $t$.

In the binomial expansion (\ref{eq:binomial_expansion}), only the terms $(N_1,N_2)$ that  are close to $(N/2,N/2)$ contribute significantly. Therefore we perform a Taylor expansion of the energy of each term at first order in $q=(N_1-N_2)/2$:
\begin{equation}
E\left(\frac{N}{2}+q,\frac{N}{2}-q \right)\approx E\left(\frac{N}{2},\frac{N}{2} \right) +(\mu_1-\mu_2) q
\label{eq:Taylor_expansion}
\end{equation}
where 
\begin{equation}
\mu_1=\left( \frac{\partial E}{\partial N_1}\right)_{N_2},\quad \mu_2=\left( \frac{\partial E}{\partial N_2}\right)_{N_1}.
\end{equation}
With this approximation, each term in the sum (\ref{eq:eq:binomial_expansion_t}) has a phase that is proportional to $(N_1-N_2)t$ and we expect a full transfer to level $|2\rangle$ after the second Ramsey pulse for :
\begin{equation}
h\nu=\mu_1-\mu_2.
\label{eq:SU2_starting_point}
\end{equation} 

\paragraph*{Validity of the expansion (\ref{eq:Taylor_expansion}).}

In order to give a necessary condition on the parameters of the problem for (\ref{eq:Taylor_expansion}) to hold, we consider the $T=0$ case and use the expression for the mean-field energy:
\begin{eqnarray}
E(N_1,N_2)&=&\frac{1}{2} (N_2-N_1)h\nu_0 + \\
&& \frac{\hbar^2}{2mL^2}\left(\tilde g_{11}N_1^2+ 2\tilde g_{12}N_1N_2 +\tilde g_{22}N_2^2\right), \nonumber
\label{eq:interaction_energy}
\end{eqnarray}
where $L^2$ is the area of the box confining the gas. One then has the exact result:
\begin{eqnarray}
E\left(\frac{N}{2}+q,\frac{N}{2}-q \right) &=& E\left(\frac{N}{2},\frac{N}{2} \right) \label{eq:quadratic_expansion_q} \\ 
&+& \left[-h\nu_0+\frac{\hbar^2}{2m}(\tilde g_{11}-\tilde g_{22})\bar n \right] q \nonumber \\
&+& \frac{\hbar^2}{2mL^2}(\tilde g_{11}+\tilde g_{22}-2\tilde g_{12})q^2. \nonumber
\end{eqnarray} 
In practice, we operate the Ramsey scheme in  the regime 
\begin{equation}
\frac{\hbar t}{2m}|\tilde g_{11}-\tilde g_{22}|\bar n \sim 1
\label{eq:regime_Ramsey_Scheme}
\end{equation} 
to obtain a good precision on the determination of $\tilde g_{11}-\tilde g_{22}$. Using the fact that for the binomial distribution, $\langle {q^2}\rangle= N/4$, we deduce that the contribution of the last line of (\ref{eq:quadratic_expansion_q}) [which was omitted in  Eq.\,(\ref{eq:Taylor_expansion})] can be neglected if:
\begin{equation}
\frac{1}{4}\left|\tilde g_{11}+\tilde g_{22}-2\tilde g_{12}\right| \lesssim |\tilde g_{11}-\tilde g_{22}|
\label{eq:inequality_1}
\end{equation}
meaning that the interspecies scattering length $a_{12}$ has to be close to the average of the intraspecies ones, $a_{11}$ and $a_{22}$. This condition is well fulfilled for $^{87}$Rb.

\subsection{Using the approximate SU(2) symmetry}

We have seen above that provided the inequality (\ref{eq:inequality_1}) is satisfied, one can achieve a full transfer from $|1\rangle$ to $|2\rangle$ in the Ramsey sequence operating in the regime (\ref{eq:regime_Ramsey_Scheme}), provided the microwave frequency is chosen such that
\begin{equation}
h\nu = \left( \frac{\partial E}{\partial N_1}\right)_{N_2}-\left( \frac{\partial E}{\partial N_2}\right)_{N_1}.
\label{eq:SU2_starting_point_2}
\end{equation}
Here, the energy $E$ is calculated for the parameters $N_1=N_2=N/2$ and the 3 scattering lengths $a_{11}$, $a_{12}$ and $a_{22}$. Suppose now that all three scattering lengths are close to each other, so that we can expand:
\begin{eqnarray}
&&E(N_1,N_2,a_{11},a_{12},a_{22})\approx E(N_1,N_2,a,a,a)+ \nonumber \\
&& \qquad \qquad (a_{12}-a_{11})\frac{\partial E}{\partial a_{12}}+(a_{22}-a_{11})\frac{\partial E}{\partial a_{22}}
\end{eqnarray}
where we have set $a\equiv a_{11}$. The SU(2) symmetry is exact at the point in parameter space where $a_{12}=a_{22}=a$.  

We note that:
\begin{equation}
\frac{\partial^2 E}{\partial N_1\, \partial a_{12}}\left(\frac{N}{2},\frac{N}{2},a,a,a \right)=\frac{\partial^2 E}{\partial N_2\, \partial a_{12}}\left(\frac{N}{2},\frac{N}{2},a,a,a \right)
\end{equation} 
so that the term $\propto(a_{12}-a_{11})$ does not contribute to (\ref{eq:SU2_starting_point_2}). This leads to 
\begin{equation}
h\,\Delta\nu =(a_{22}-a_{11})\,\left[ \frac{\partial^2 E}{\partial N_1 \,\partial a_{22}}-\frac{\partial^2 E}{\partial N_2 \,\partial a_{22}} \right]\left(\frac{N}{2},\frac{N}{2},a,a,a \right).
\end{equation}

Now, the Hamiltonian of the binary system for a regularized zero-range potential is
\begin{equation}
\hat H=\hat H_0+\sum_{i,j}a_{ij}\hat K_{ij}
\end{equation}
where 
\begin{equation}
\hat K_{ij}=\frac{2\pi \hbar^2}{m}\int\hskip-3mm\int \hat \psi_i^\dagger(\bs r)\, \hat \psi_j^\dagger(\bs r')\,\hat \delta(\bs r-\bs r')\,\hat \psi_j(\bs r')\,\hat \psi_i(\bs r)\ \D^3 r\;\D^3 r'.
\end{equation}
 Hellmann--Feynman theorem thus leads to:
\begin{equation}
h\,\Delta \nu \approx (a_{22}-a_{11})\,\left[\frac{\partial \langle \hat K_{22}\rangle}{\partial N_1}-\frac{\partial \langle \hat K_{22}\rangle}{\partial N_2}\right]  \left(\frac{N}{2},\frac{N}{2},a,a,a \right)
\end{equation}
At the SU(2) point, we can connect the two-component system with the single component system with the same scattering length:
\begin{equation}
\langle \hat K_{22}\rangle=\frac{N_2^2}{(N_1+N_2)^2}\langle \hat K\rangle
\label{eq:hypothesis_SU2}
\end{equation}
We then find:
\begin{equation}
N\,h\,\Delta \nu \approx  (a_{22}-a_{11})\,\langle \hat K\rangle,
\label{eq:SU_2_result}
\end{equation}
which also reads, setting $\Delta a=a_{22}-a_{11}$:
\begin{equation}
C=\frac{16\pi^2 m a^2 N}{\hbar} \,\frac{\Delta \nu}{\Delta a},
\end{equation}
and which coincides with the expressions (3,5) of the main text.


\section{Contact and two-body correlation within Bogoliubov approach} 

\subsection{Bogoliubov operators and contact}

We consider a 2D Bose gas confined in a square box $L\times L$ with periodic boundary conditions. We denote $\hat a_{\bs k}$ the operator that annihilates a particle with momentum $\hbar \bs k$. We assume that the temperature is low enough so that most of the particles accumulate in the ground state of the box $\bs k=0$. Since the confining box has a finite size, this does not violate Mermin-Wagner theorem, which holds for a gas in the thermodynamic limit. Note that instead of assuming a macroscopic population of $\bs k=0$, one may also use another version of the Bogoliubov approach in terms of phase and density fluctuations (see e.g. \cite{Mora:2003}). In that approach, which leads to the same results as the one used here, one assumes that the density fluctuations are small and that the phase fluctuations can be expanded as a Fourier series (no isolated vortex).

The Bogoliubov Hamiltonian is diagonalized by introducing the bosonic operators
$\hat b_{\bs k}=u_k \hat a_{\bs k}-v_k\hat a_{-\bs k}^\dagger$ %
with
\begin{equation}
u_k,v_k=\pm\left[\frac{k^2 + 2 \tilde g \bar n}{2k(k^2+4\tilde g \bar n)^{1/2}}\pm \frac{1}{2}  \right]^{1/2}, 
\label{eq:u_v_k_2D}
\end{equation}
and the energy of the Bogoliubov modes
\begin{equation}
\epsilon_k=\frac{\hbar^2k}{2m}\left[k^2+4\tilde g \bar n \right]^{1/2}.
\end{equation}
The Bogoliubov Hamiltonian reads:
\begin{equation}
\hat H=E_0+\sum_{\bs k} \epsilon_k\;\hat b_{\bs k}^\dagger \hat b_{\bs k}.
\end{equation}
In the case studied in the paper, where the thickness $a_z$ of the gas is large compared to the scattering length, the ground-state energy $E_0$ can be estimated by averaging the mean-field 3D result: 
\begin{equation}
E_0^{\rm (3D)}=\frac{2\pi \hbar^2 a}{m}n^{\rm (3D)}N 
\end{equation}
over the Gaussian density profile $n^{\rm (3D)}(z)=\bar n\; \E^{-z^2/a_z^2}/\sqrt{\pi a_z^2}$ along the $z$ direction:
\begin{equation}
E_0=\frac{\int E_0^{\rm (3D)}(z)\ n(z)\ \D z}{\int n(z)\ \D z}=\frac{\hbar^2 \tilde g}{2m} \bar n N
\end{equation} 

The thermal averages are $\langle \hat b_{\bs k}\rangle=0$ and
$\langle \hat b_{\bs k}^{\dagger} \hat b_{\bs k'}\rangle=\delta_{\bs k,\bs k'}\,{\cal N}_k$, 
where ${\cal N}_k$ is the Bose--Einstein distribution
\begin{equation}
{\cal N}_k= \left[\E^{\epsilon_k/\kB T}-1  \right]^{-1}.
\end{equation}
The internal energy in thermal equilibrium thus reads:
\begin{equation}
E=E_0+\sum_{\bs k} \epsilon_k {\cal N}_k.
\end{equation}
The contact is by definition proportional to the derivative of this energy with respect to $a$ at constant entropy, i.e. at constant populations ${\cal N}_k$ of the modes, which gives:
\begin{equation}
C=C^{T=0}+C^{\rm thermal}
\end{equation}
with
\begin{equation}
C^{T=0}=\frac{8\pi m a^2}{\hbar^2}\; \frac{\partial E_0}{\partial a}=C_0
\end{equation}
and
\begin{eqnarray}
C^{\rm thermal}&=&\frac{8\pi m a^2}{\hbar^2}\left[ \sum_{\bs k} \frac{\partial \epsilon_k}{\partial a}\, {\cal N}_k\right]\nonumber \\
&=& C_0\ \frac{2}{N} \sum_{\bs k}\frac{k}{\sqrt{k^2+4\tilde g \bar n}}\, {\cal N}_k.
\end{eqnarray}


\subsection{Density fluctuations}

\paragraph{Average density.}

The average density of the gas is calculated from $
\bar n=\langle \hat n(\bs r)\rangle$ with  $\hat n(\bs r)=\hat \psi^\dagger(\bs r)\hat \psi(\bs r)$,
and it can be split into a $T=0$ and a thermal component: 
\begin{equation}
\bar n^{T=0}=
\frac{N_0}{L^2}
+\frac{1}{L^2}\sum_{\bs k\neq 0}v_k^2
\end{equation}
and 
\begin{equation}
\bar n^{\rm thermal}=\frac{1}{L^2}\sum_{\bs k\neq 0} (u_k^2+v_k^2) \;{\cal N}_k.
\end{equation}


\paragraph{Density correlations.} 

We start from the 4-field correlation function written in normal order
$G_2(\bs r)=\langle \hat \psi^\dagger(0)\hat \psi^\dagger(\bs r) \hat \psi(\bs r)\hat \psi(0) \rangle$,
which we expand up to first order in $n^{\rm thermal}/\bar n$:
\begin{eqnarray}
&&G_2(\bs r)=\frac{N_0^2}{L^4}\ +\ \frac{N_0}{L^4} \times \\
& & 
\sum_{\bs k\neq 0}\E^{\I \bs k\cdot \bs r} \left( \langle \hat a_{-\bs k}\hat a_{\bs k}\rangle + \langle \hat a_{-\bs k}^\dagger\hat a_{\bs k}^\dagger\rangle
+2 \langle \hat a_{\bs k}^\dagger \hat a_{\bs k}\rangle \right) 
+ 2 \langle \hat a_{\bs k}^\dagger \hat a_{\bs k}\rangle 
\nonumber 
\end{eqnarray}
We can then calculate the $g_2$ function used in the main text:
\begin{equation}
g_2(\bs r)=\frac{G_2(\bs r)}{\bar n^2} =g_2^{T=0}(\bs r)+g_2^{\rm thermal}(\bs r)
\end{equation}
and we find by at first order in $(\bar n-\bar n_0)/\bar n$ [see e.g. \cite{Mora:2003}]:
\begin{equation}
g_2^{T=0}(\bs r)=1+\frac{2}{N}\sum_{\bs k\neq 0}\E^{\I \bs k\cdot \bs r} v_k(u_k+v_k)
\label{eq:g2_T_0_Bogo}
\end{equation}
and
\begin{equation}
g_2^{\rm thermal}(\bs r)= \frac{2}{N_0}\sum_{\bs k\neq 0}\E^{\I \bs k\cdot \bs r} \left( u_k+v_k\right)^2\;{\cal N}_k. \label{eq:thermal_g2}
\end{equation}
We notice that
\begin{equation}
(u_k+v_k)^2=\frac{k}{\sqrt{k^2+4\tilde g \bar n}}
\end{equation}
which shows the relation (6) of the main text:
\begin{equation}
\frac{C^{\rm thermal}}{C_0}=g_2^{\rm thermal}(0).
\end{equation}
On the other hand, the integral giving $g_2^{T=0}$ in $\bs r=0$ is UV divergent since $v_k\propto 1/k^2$ and $u_k+v_k\sim 1$ at infinity. 


\subsection{Lee-Huang-Yang (LHY) correction \cite{Lee:1957}}

In 3D and at zero-temperature, the first beyond-mean-field correction to the contact is (see e.g. Eq.(2) in \cite{wild2012measurements})
\begin{equation}
\frac{\delta C}{C}=\frac{5}{2}\times\frac{128}{15\sqrt{\pi}}\sqrt{n^{\rm (3D)}a^3}.
\end{equation}
In our setup, the average 3D density is $n^{\rm (3D)}=\bar n/(a_z\sqrt {2\pi})$. For a 2D density $\bar n=40$\,atoms/\si{\micro}m$^2$ and $a_z=160$\,nm, this gives $\bar n^{\rm (3D)}\approx 1.0\times10^{14}$\,atoms/cm$^3$ and $\delta C/C\approx 4.7\,\%$. For this $\bar n$, the mean-field contribution to the contact corresponds to a shift $\Delta  \nu=-22$\,Hz (Fig.\,2 of the main text), and the LHY correction is $\approx 1\,$Hz, within the uncertainty of our measurements. Note that a more precise theoretical estimate of the LHY correction for our planar geometry should start from the general expression of the ground-state energy of a 2D Bose gas \cite{Schick:1971,Mora:2009_PhysRevLett.102.180404,Fournais:2019} and the relation between the 2D scattering length and the 3D one \cite{Petrov:2001}.


\subsection{Estimate for the contribution of the 3-body contact}

Using the transition rates derived in \cite{Braaten:2011_PhysRevLett.106.153005}, Fletcher et al.\;\cite{Fletcher:2017} have shown that the contribution of the 3-body contact to the many-body resonance shift  is related to the shift due to the 2-body contact by:
\begin{equation}
\frac{\Delta \nu_3}{\Delta \nu_2}=5.0\pi^2\,a\;\frac{C_3}{C_2}.
\end{equation}
Now an estimate of $C_3/C_2$ for a dilute BEC is provided by \cite{Smith:2014_PhysRevLett.112.110402}:
\begin{equation}
\frac{C_3}{C_2}\sim 0.02\;n^{\rm (3D)}a^2 
\end{equation}
so that the contribution of the 3-body contact is reduced by a factor $\sim n^{\rm (3D)}a^3\sim10^{-5}$ with respect to the contribution of the 2-body contact. Even though the 2D nature of the thermodynamics of our gas may bring some significant corrections to this crude estimate, we can safely assume that effects related to the 3-body contact cannot be detected with our experimental protocol.

\section{Virial expansion for a 2D Bose gas}

In \cite{ren2004virial}, H.c Ren gives the result of perturbative thermodynamics applied to a regularized contact potential in 2D. Strictly speaking, this is not a virial expansion, i.e. an expansion in powers of  density or  fugacity, since the author takes exactly into account all powers of $\bar n$ in the ideal gas case.

Starting from the 2D scattering length $a_2$, Ren introduces the dimensionless coupling 
\begin{equation}
\alpha(T)=\frac{1}{\ln\left( \frac{\lambda^2}{2\pi a_2^2} \right)+\gamma} 
\label{eq:def_alpha_Ren}
\end{equation}
where $\lambda(T)$ is the thermal wavelength and $\gamma$ the Euler constant, which is related to $\tilde g$ by  $\tilde g \approx 4\pi \alpha$. He then performs a systematic expansion of various thermodynamic functions in powers of $\alpha$. Note that the $T$ dependence of $\alpha$ explicitly breaks the scale invariance of the problem, as expected after regularization of the contact interaction in 2D. However for our experimental parameters, this $T$-dependence plays a negligible role. 

The value of the free energy $F$ reads at order 2 in $\alpha$:
\begin{eqnarray}
F(N,L^2,T,a_2)&=&F_0(N,L^2,T)\ + \alpha\,\frac{\ 4\pi\hbar^2N^2}{mL^2} \nonumber \\
&-&\ \alpha^2\,\frac{8\pi L^2\hbar^2}{m\lambda^4} \phi\left[ 1-\E^{-N\lambda^2/L^2}\right],
\label{eq:F_vs_dens}
\end{eqnarray}
where $F_0$ is the ideal Bose gas result and where the function $\phi(z)$ is defined by:
\begin{equation}
\phi(z)=B(z)+\frac{1}{2}D(z)
\end{equation}
with
\begin{equation}
B(z)=\sum_{r,s,t=1}^{\infty}\frac{z^{r+s+t}}{\sqrt{rs(r+t)(s+t)}}
\ln \frac{\sqrt{(r+t)(s+t)}+\sqrt{rs}}{\sqrt{(r+t)(s+t)}-\sqrt{rs}}
\end{equation}
and 
\begin{equation}
D(z)=\sum_{r,s=1}^{\infty}\frac{z^{r+s}}{rs} \ln\frac{2rs}{r+s}.
\end{equation}

Tan's contact 
\begin{equation}
C=\frac{8\pi m a^2 }{\hbar^2}\left( \frac{\partial F}{\partial a}\right)_{N,L^2,T}
\end{equation}
can then be calculated using (\ref{eq:def_alpha_Ren}) together with the link between the 2D ($a_2$) and 3D ($a$) scattering lengths and the size of the ground state along the $z$ direction ($a_z$) \cite{Petrov:2001,Pricoupenko:2007}
\begin{equation}
a_2 \approx 2.092\, a_{z} \exp \left( - \sqrt{\frac{\pi}{2}}\frac{a_z}{a} \right).
\label{eq:a2_vs_a}
\end{equation}
The result is plotted in Fig.\,3 of the article.

\bibliographystyle{apsrev}
\bibliography{contact}

\end{document}